\documentclass[11pt,twoside]{article}
\usepackage{asp2010}

\resetcounters

\begin{document}

\title{Increasing supercycle lengths of active SU~UMa-type dwarf novae}
\author{Magdalena~Otulakowska-Hypka$^1$ and Arkadiusz~Olech$^1$}
\affil{$^1$Nicolaus Copernicus Astronomical Center of the Polish Academy of Sciences, Warsaw, Poland}

\begin{abstract}
We present observational evidence that supercycle lengths of the most active SU~UMa-type stars are increasing during the past decades. We analysed a large number of photometric measurements from available archives and found that this effect is generic for this class of stars, independently of their evolutionary status. This finding is in agreement with previous predictions and the most recent work of \cite{2012Patterson} on BK~Lyn.
\end{abstract}

\section{Introduction}

\textit{SU~UMa-type dwarf novae} 
are objects with a white dwarf as the primary and low mass main-sequence star as the secondary. Mass from the secondary is flowing through the accretion disk onto the white dwarf. This is the region where characteristic outbursts take place.
The main feature of this class of stars is the fact that beside normal outbursts they also show superoutbursts, which are less frequent but brighter and last about ten times longer than normal outbursts. 

\begin{center}
\begin{figure}[h]
\includegraphics[width=0.9\textwidth]{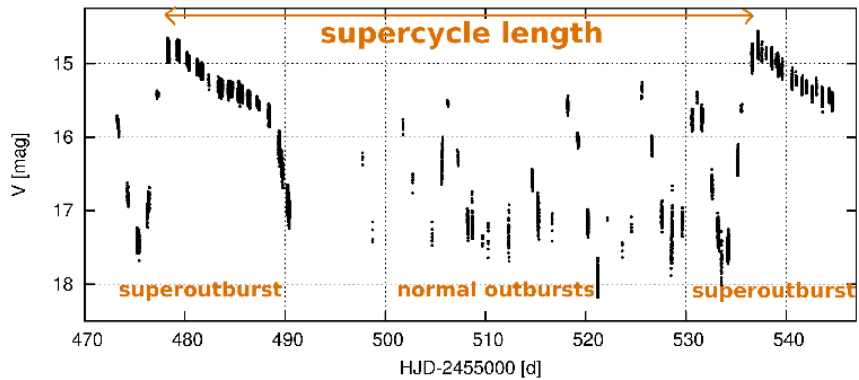}
\label{fig:lc}
\caption{Example light curve of an active SU UMa star.}
\end{figure}
\end{center}

\textit{Supercycle length}, $\rm{P_{sc}}$, is the period between two successive superoutbursts. It is one of the most fundamental properties of SU~UMa stars and is specific for each of them (Fig.~\ref{fig:lc}).

In our recent work \citep{2012MOH} we surprisingly found that the $\rm{P_{sc}}$ of IX~Draconis is increasing with a constant rate since the last twenty years. We decided to investigate the same issue for other active SU~UMa stars.

\section{Objects}

\begin{figure}[h]
\includegraphics[width=0.6\textwidth,angle=270]{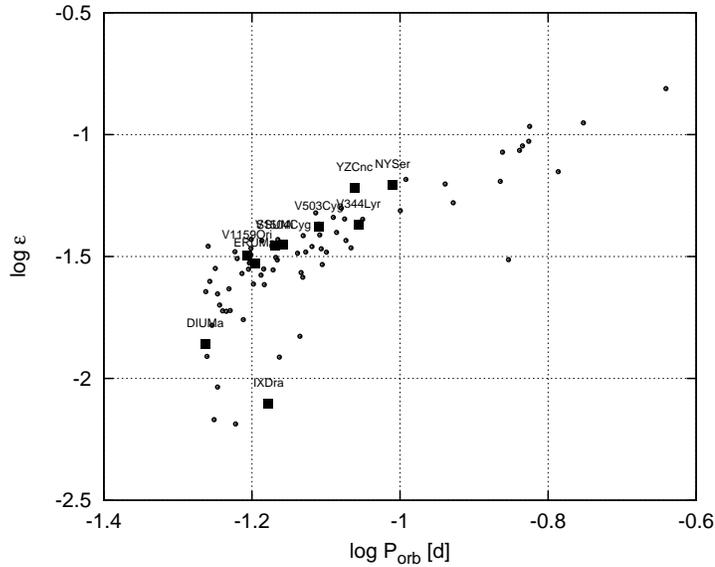}
\label{fig:epsilon}
\caption{Relation between the orbital period-superhump period excess ($\epsilon$) and the orbital period ($P_{orb}$) for dwarf novae stars, which can be used to trace the evolution of these objects.}
\end{figure}

For our analysis we selected the most active objects ($\rm{P_{sc}}$ $<$ 120~days) of this class of stars.
They are diverse in terms of evolutionary status.
We marked them with squares in Fig.~\ref{fig:epsilon}. 

\section{Photometric data}
   
We searched all accessible professional and amateur photometric databases to create joint light curves which are as complete as possible. We used data from the following archives: 
the ASAS\footnote{http://www.astrouw.edu.pl/asas/} project \citep{1997Pojmanski}, 
the MEDUZA\footnote{http://var2.astro.cz/} project of the Variable Star and Exoplanet Section of the Czech Astronomical Society \citep{2010meduza},
and the databases with amateur observations: 
AAVSO\footnote{http://www.aavso.org/}, 
AFOEV\footnote{http://cdsarc.u-strasbg.fr/afoev/},
and BAAVSS\footnote{http://www.britastro.org/vss/}.

What is more, we had some extra data mostly from our previous observational campaigns, which was also used in this work. This data comes from: 
\cite{1995Kato},
\cite{2004Olech},
\cite{2006Olech},
\cite{2008Olech}, 
\cite{2009Rutkowski}, and
\cite{2012MOH}.

\section{Results}

\begin{figure}[h]
\includegraphics[angle=270,width=0.43\textwidth]{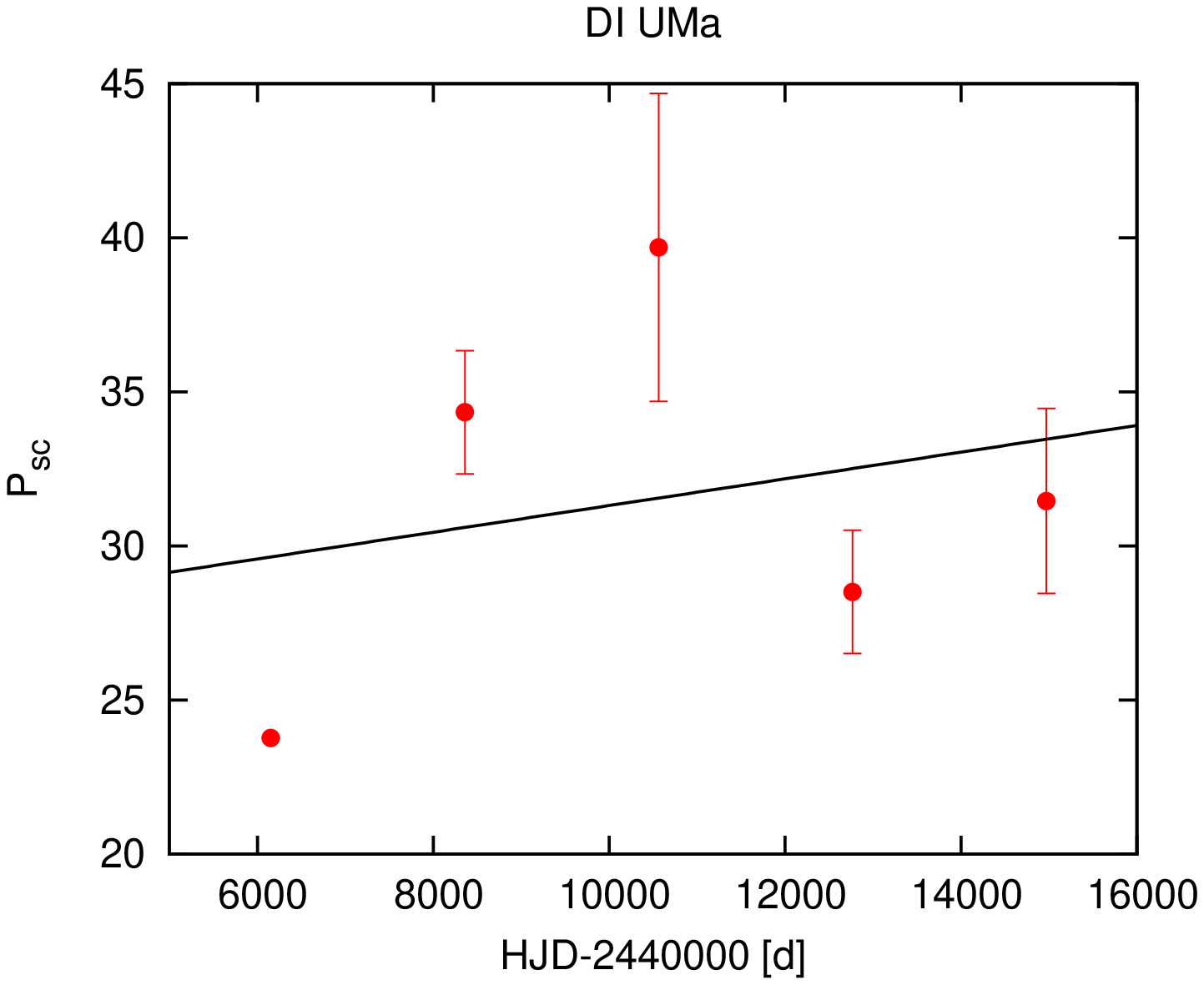}
\includegraphics[angle=270,width=0.43\textwidth]{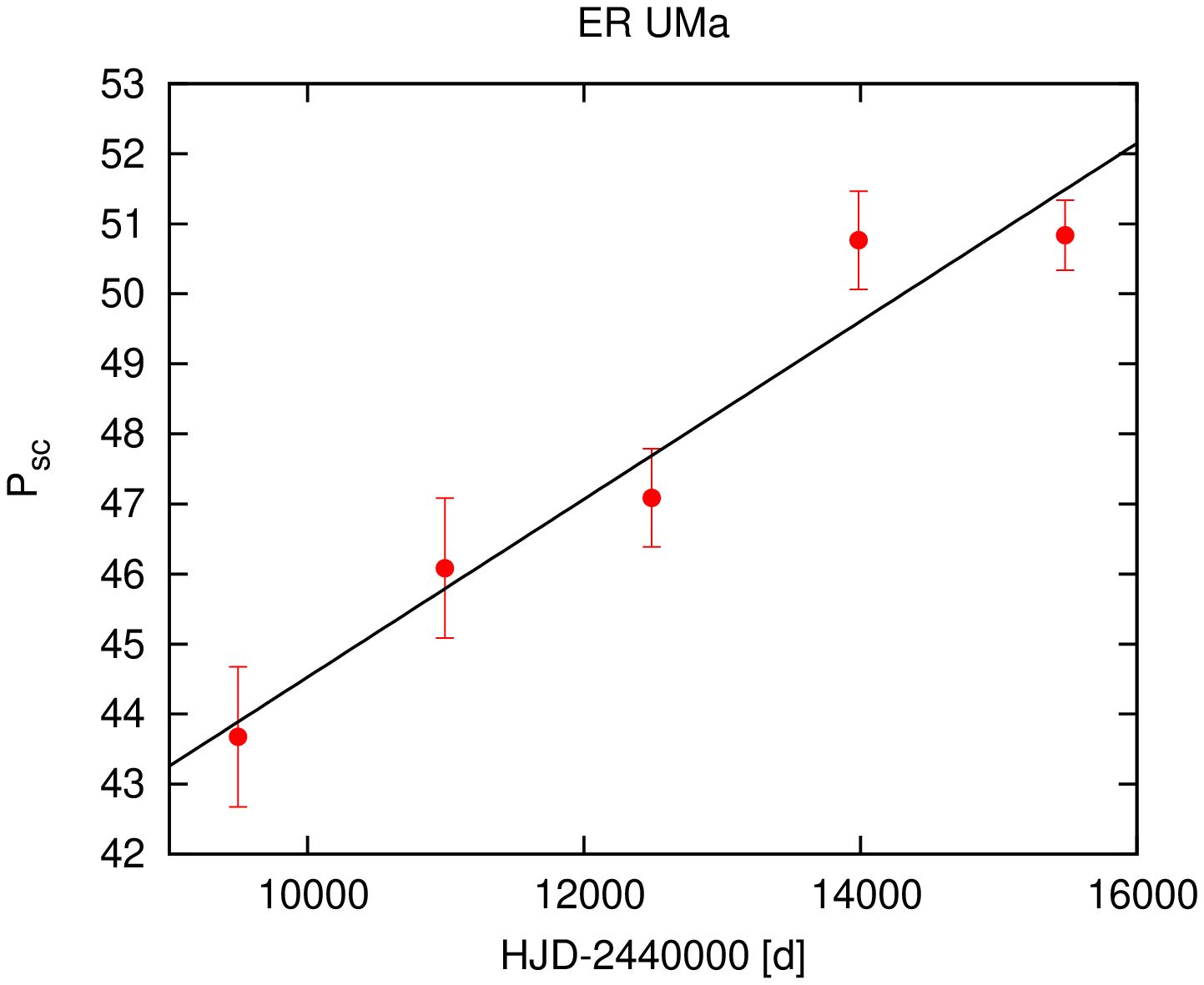}\\
\includegraphics[angle=270,width=0.43\textwidth]{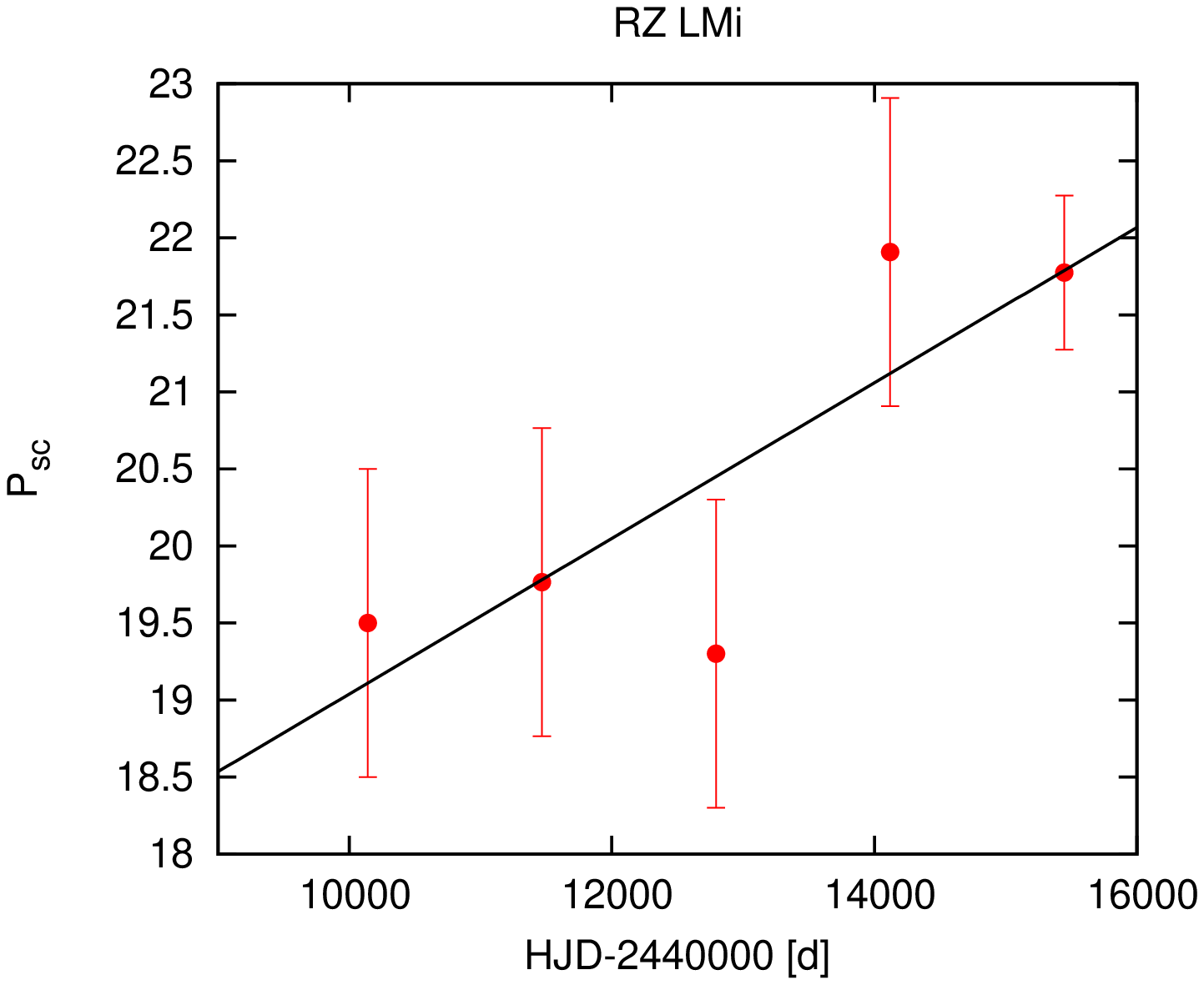}
\includegraphics[angle=270,width=0.43\textwidth]{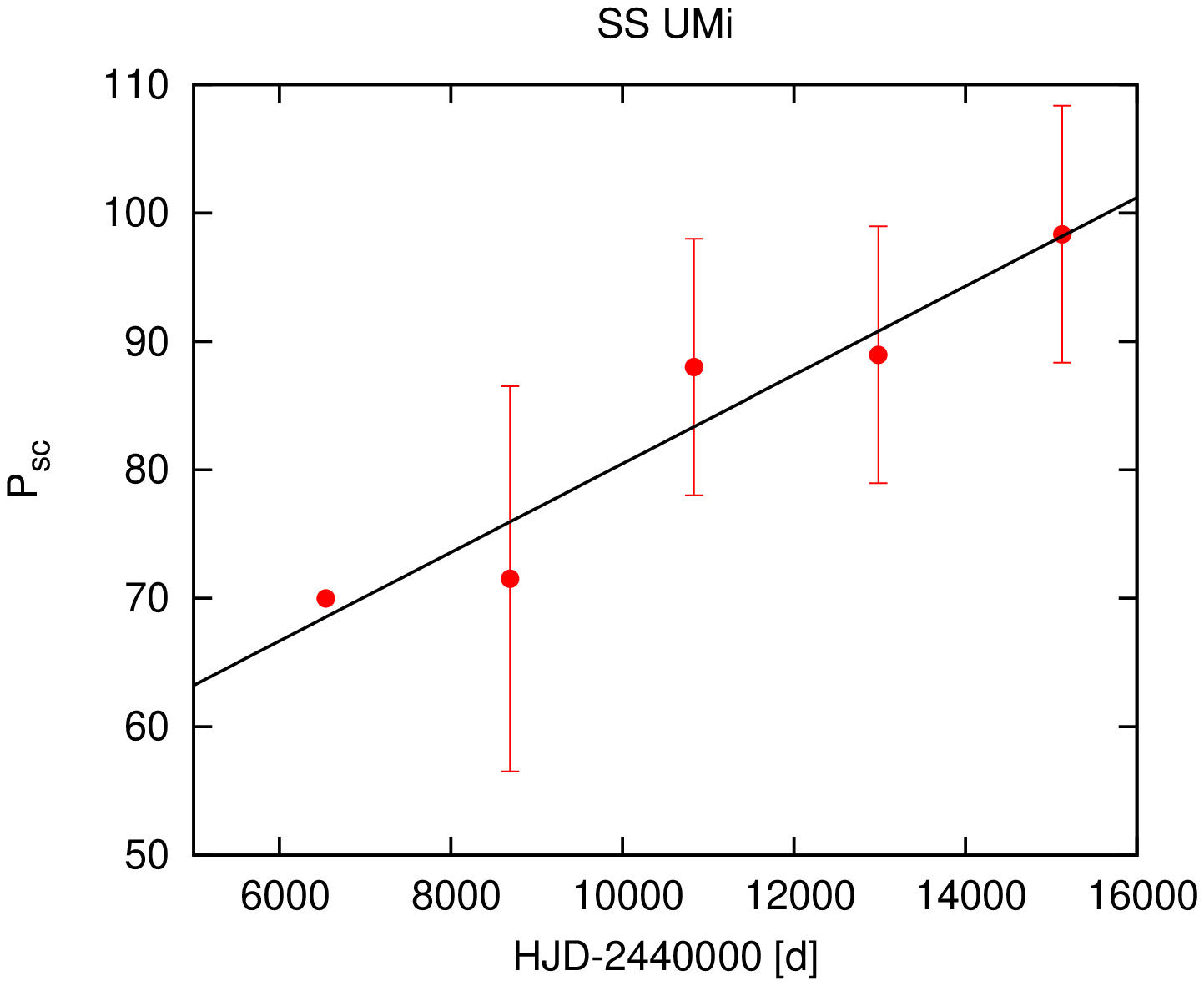}
\caption{Examples of results}
\label{fig:psc}
\end{figure}

So far, all of the analysed objects have positive values of their period derivatives (Tab.~\ref{tab-results}). For all of them the $P_{sc}$ is increasing (Fig.~\ref{fig:psc}). 
There are some subtle fluctuations for short time scales, but the general trend is the same in each case.

\begin{center}
\begin{table}[h]
\caption{Derived rates of changes of $P_{sc}$}
\begin{tabular}{c|c}
Object		& 	$\dot{P_{sc}}$ \\
\hline
DI~UMa 			&	$4.3 \times 10^{-4}$ \\
ER~UMa 			&	$1.3 \times 10^{-3}$ \\
IX Dra	 		&	analysis in progress	 \\
NY Ser			&	analysis in progress	 \\
RZ~LMi 			&	$5.0 \times 10^{-4}$ \\
SS~UMi 			&	$3.5 \times 10^{-3}$ \\
V1159~Ori 		&	$1.1 \times 10^{-3}$ \\
V1504 Cyg		&	analysis in progress	 \\
V344 Lyr		&	analysis in progress	 \\
V503~Cyg 		&	$6.4 \times 10^{-4}$ \\
YZ~Cnc 			&	$3.0 \times 10^{-3}$ \\
\end{tabular}
\label{tab-results}
\end{table}
\end{center}

Increasing supercycle lengths mean that the mass transfer rates are decreasing for these objects over last decades. This is in agreement with the scenario of the evolution of BK~Lyn presented by \cite{2012Patterson}, which seems to be a general case.
This phenomenon is important in the context of evolution of such systems.

\acknowledgements 
We acknowledge with thanks the variable star observations 
from the AAVSO International Database, operated in USA,
the AFOEV database, operated at CDS, France,
the BAAVSS database, operated in UK,
and the Variable Star and Exoplanet Section of the Czech Astronomical Society, 
contributed by observers worldwide and used in this research.
MOH acknowledges support provided by the Organizers which allowed her to participate in the conference and the Polish National Science Center grant awarded by decision number DEC-2011/03/N/ST9/03289.

\bibliography{Psc_bib}

\end{document}